\documentclass[prc,aps,twocolumn,showpacs]{revtex4}
\usepackage{graphicx}
\begin{document}

\title{Effect of double-polarization data in \\
       fits to single-pion photoproduction}

\author{%
Richard A. Arndt}
\affiliation{Center for Nuclear Studies, Department of
Physics,\\
The George Washington University, Washington DC 20052}
\author{%
Igor I. Strakovsky}
\affiliation{Center for Nuclear Studies, Department of
Physics,\\
The George Washington University, Washington DC 20052}
\author{%
Ron L. Workman}
\affiliation{Center for Nuclear Studies, Department of
Physics,\\
The George Washington University, Washington DC 20052}

\begin{abstract}
We investigate the influence of new Jefferson Lab
polarization-transfer, $C_{x'}$ and $C_{z'}$, and recoil 
polarization, $P_y$, measurements in multipole fits to 
the single-pion photoproduction database. Results are 
compared to those found with the addition of a 
double-polarization quantity associated with the 
Gerasimov$-$Drell-Hearn sum rule.
\end{abstract}
\vspace*{0.5in}


\maketitle

The single-pion photoproduction process has been studied
extensively, both experimentally and theoretically, for   
several decades. Over the resonance region, there now
exists a database of approximately twenty thousand
angular and total cross section measurements. This set
consists mainly of unpolarized cross sections, many of
these having been obtained without the benefit of tagged 
photons.

Experimental programs now underway have been adding very 
precise cross sections, many involving tagged photons, 
polarization, and a wide angular acceptance.  Coverage 
is not yet complete. As a result, our knowledge of the  
underlying resonance contributions is based on a mix of 
data, old and new, some of which are certainly flawed. 
Fits to the full data set tend to favor the newer 
measurements~\cite{GW_pi}.  However, experience has 
shown that sensitive quantities can be biased by older 
results, particularly if they cover a gap in the newer 
measurements, and are presented with comparable 
uncertainties~\cite{bonn_e2}.

Polarization quantities have been particularly influential
in constraining multipole analyses.  For this process, 
there are 8 independent quantities, each providing a new 
constraint on the 4 helicity amplitudes.  In the $\Delta 
(1232)$ region, a single resonance is dominant, and in 
this case (only) each new quantity can be translated into 
a constraint on a single ($M_{1+}^{3/2}$) multipole and 
its interference with other smaller multipoles.  Here, and 
also at higher energies, the beam-polarization ($\Sigma$) 
has been useful in revealing problems~\cite{graal}
with the existing SAID~\cite{GW_pi} and MAID~\cite{maid}
fits.

Until recently, very few double-polarization measurements  
existed, and the available data had large uncertainties,
weakening their influence in unweighted fits.  As a
by-product of the ongoing GDH measurements, angular 
distributions of the difference of helicity 1/2 and 3/2 
cross sections, related to the beam-target polarization 
quantity ($E$)~\cite{bds}, have been measured and fitted~
\cite{GDH}.  While SAID and MAID gave qualitatively 
correct predictions for this quantity, deviations 
starting at 700~MeV were linked to the $N(1520)$ 
photo-decay amplitudes~\cite{GDH}.

A double-polarization experiment~\cite{gilman}, giving 
the new beam-recoil quantities $C_{x'}$ and $C_{z'}$, 
has been completed by the Hall~A Collaboration at 
Jefferson Lab.  These measurements, which extend from 
0.83~GeV to 4.03~GeV, are given with high precision, and 
cover a region dominated by older data.  While SAID and 
MAID predict the lowest energy results, at energies of 
1~GeV and beyond even qualitative agreement is lost~
\cite{sabit}.  In addition, just beyond the energy range 
of SAID, the recoil polarization ($P_y$), also measured 
in Ref.~\cite{gilman}, appears to have a rapid angular 
variation not seen below 2~GeV.

\begin{figure*}
\centering{
\includegraphics[height=0.6\textwidth, angle=90]{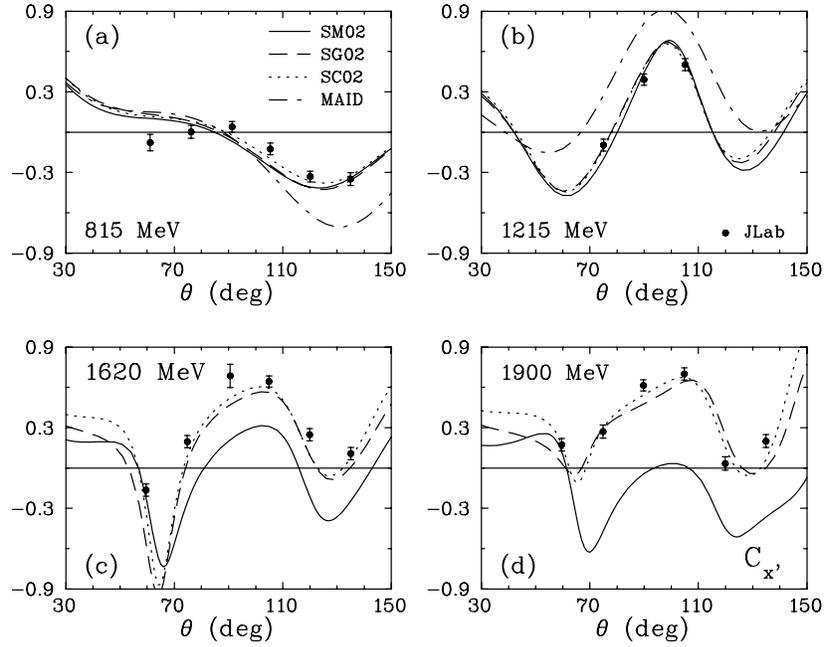}
}\caption{Polarization transfer $C_{x'}$ in neutral pion
          photoproduction at medium angles in the fixed
          lab frame: (a) E$_{\gamma}$ = 815~MeV, (b) 1215~MeV,
          (c) 1620~MeV, and (d) 1900~MeV.  Solid (dashed-dot)
          line gives the SM02~\protect\cite{GW_pi} (MAID~
          \protect\cite{maid}) solution.  Dashed (dotted) 
          line shows the standard SG02 (forced SC02) solution 
          including the data of Ref.~\protect\cite{gilman}.
          Plots (a) and (b) have centered angular points 
          within a 18~MeV bin; plots (c) and (d) within a 
          45~MeV bin.} \label{fig:fig1}
\end{figure*}

\begin{figure*}
\centering{
\includegraphics[height=0.6\textwidth, angle=90]{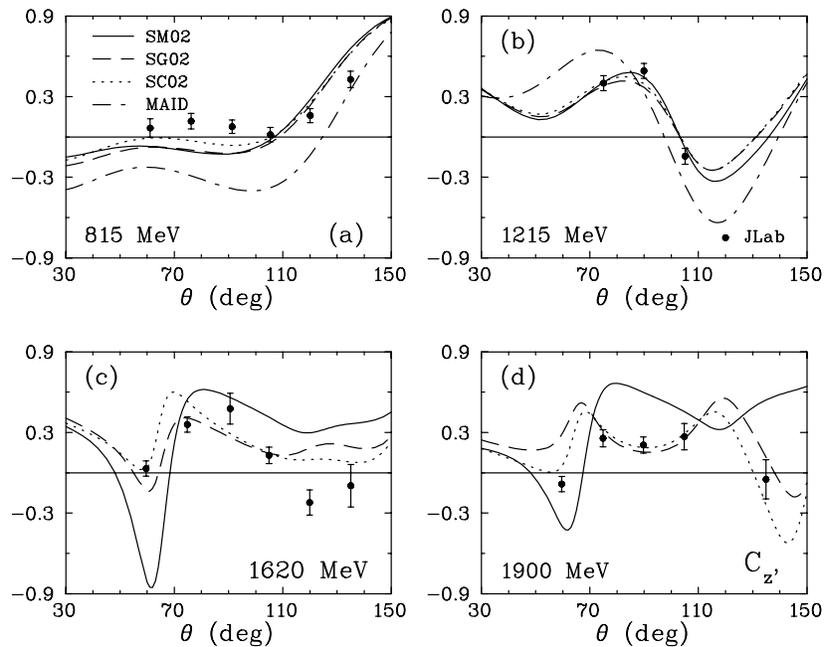}
}\caption{Polarization transfer $C_{z'}$ in neutral pion
          photoproduction at medium angles in the fixed
          lab frame.  The notation is the same as for
          Fig.~\ref{fig:fig1}.} \label{fig:fig2}
\end{figure*}

We have attempted to fit these new values of $P_y$, 
$C_{x'}$, and $C_{z'}$ in order to determine their effect 
on our multipole solution.   In Figs.~\ref{fig:fig1} and
\ref{fig:fig2}, we compare our results (SM02) prior to the 
measurement of Ref.~\cite{gilman}, (SG02) including the 
new $P_y$, $C_{x'}$, and $C_{z'}$ points, and (SC02) a 
forced fit.  At 815~MeV and 1215~MeV, comparisons with 
MAID are also presented.  The forced fit has weighted
the data of Ref.\cite{gilman} by a factor of 4.  By   
weighting data, we magnify changes in the multipole
amplitudes, and more clearly see where data conflicts 
occur.  In the case of the E2/M1 ratio, for example, an 
early forced fit~\cite{forced} to $\Sigma$ data 
resulted in a value of $-2.9\%$, when the predominant 
value was $-1.5\%$.

\begin{figure*}
\centering{   
\includegraphics[height=0.6\textwidth, angle=90]{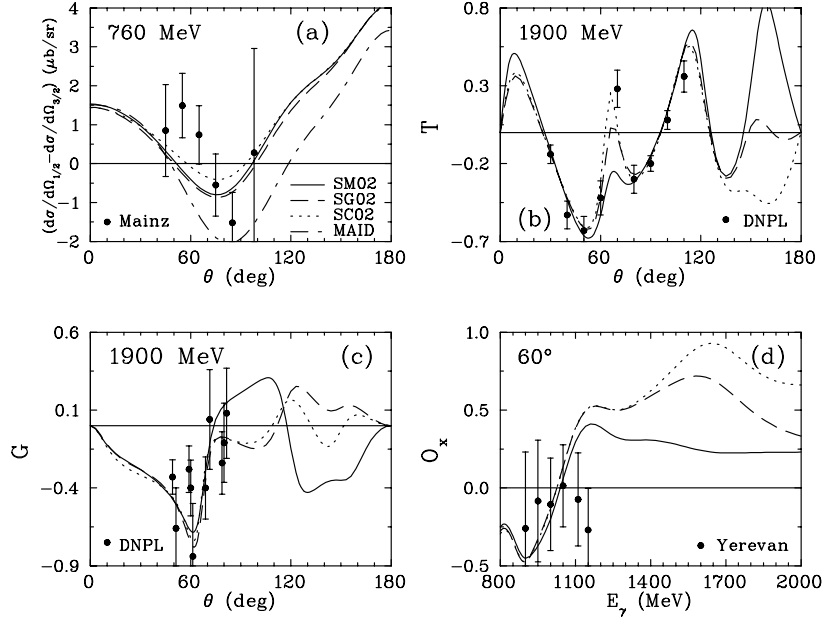} 
}\caption{Polarized measurements in neutral pion
          photoproduction: (a) $d\sigma /d\Omega_{1/2}- 
          d\sigma /d\Omega_{3/2}$, (b) T, (c) G, and (d)
          $O_x$.  The notation for curves is the same as
          for Fig.~\ref{fig:fig1}.  Data for (a) are from 
          Mainz~\protect\cite{GDH}, data for (b) and (c)
          are from DNPL~\protect\cite{barbour,bs79}, and 
          data for (d) are from Yerevan~\protect\cite{yer}.} \label{fig:fig3}
\end{figure*}

In Figs.~\ref{fig:fig1} and \ref{fig:fig2}, it is clear that 
both SAID and MAID, at the lower energies where they can be 
compared, have a shape consistent with the $C_{x',z'}$ data. 
The deviations at 1620~MeV and 1900~MeV are much larger. 
However, except for a few points, the fit including data 
from Ref.~\cite{gilman} is not appreciably different from 
the ``forced" fit, and appears to adequately describe these 
new results.  In Fig.~\ref{fig:fig3}, we show that forcing 
a fit to this new set has not degraded our fit to the GDH 
measurements.  Other double-polarization observables 
deviate considerably, according to the weighting of these 
polarization-transfer data.  These revised fits have been 
carried out using the parametrization described in Ref.~
\cite{GW_pi}.  No increase in the angular momentum limit 
for parametrized waves was found to be necessary.

\begin{figure*}
\centering{
\includegraphics[height=0.6\textwidth, angle=90]{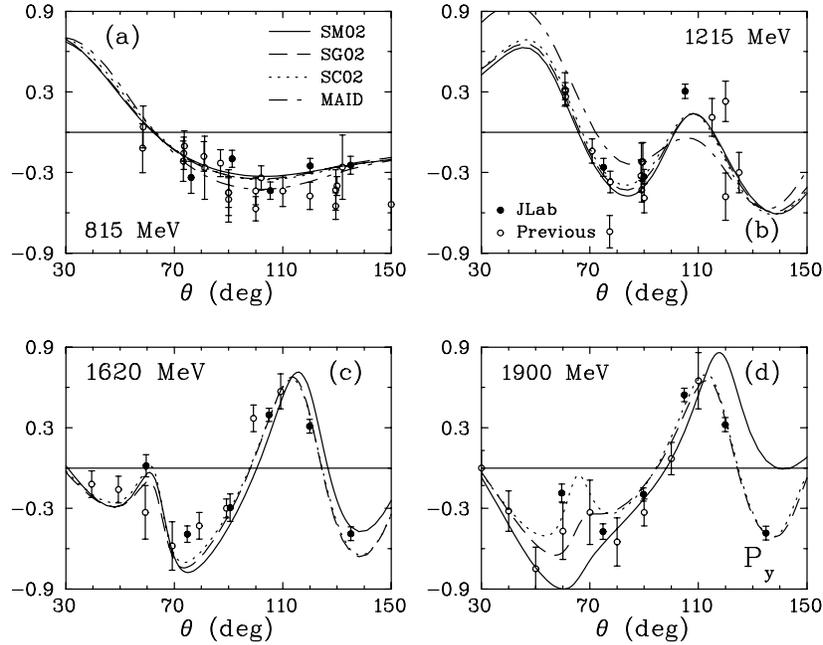}   
}\caption{Induced polarization $P_{y}$ in neutral pion
          photoproduction at medium angles.  The notation 
          for curves is the same as for Fig.~\ref{fig:fig1}.  
          Previous Bonn, DNPL, Frascati, Kharkov, SLAC, 
          Tokyo, and Yerevan measurements are available 
          in the SAID database~\protect\cite{said}.
          (a) 815~MeV plot covers 799 to 825~MeV range,
          (b) 1205~MeV plot covers 1192 to 1225~MeV
          range, (c) 1620~MeV plot covers 1600 to
          1639~MeV range, and (d) 1900~MeV plot covers   
          1876 to 1921~MeV range.} \label{fig:fig4}
\end{figure*}

In Fig.~\ref{fig:fig4}, we display the above fits for 
$P_y$.  As more weight is given to the measurements of Ref.~
\cite{gilman}, an interesting feature emerges in the
angular distribution at 1900~MeV.  A sharp structure at
about 60$^{\circ}$ is seen in the forced fit.  This is   
correlated with the improved fit seen in Fig.~
\ref{fig:fig2}(d), for $C_{z'}$, and Fig.~\ref{fig:fig3}(b), 
for the target-polarization quantity $T$.  This feature of 
the single-spin asymmetries is not new, however, and was 
present in the old analysis of Barbour, Crawford, and 
Parsons~\cite{barbour}.

\begin{figure*}
\centering{
\includegraphics[height=0.6\textwidth, angle=90]{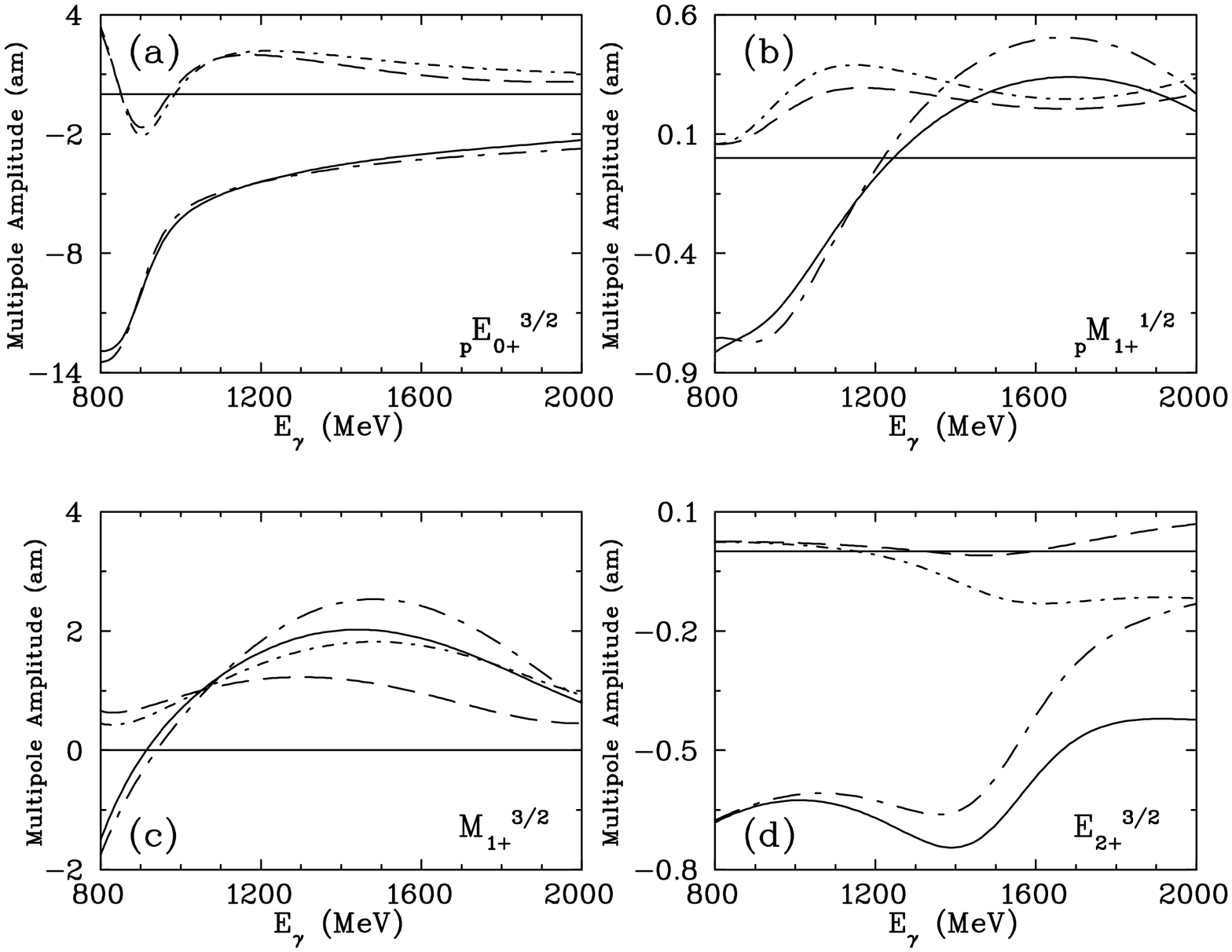}
}\caption{Selected multipole amplitudes.    
          Solid (dashed) curves give the real (imaginary) 
          parts of amplitudes corresponding to the SM02  
          ~\protect\cite{GW_pi} solution.  The real  
          (imaginary) parts of the forced SC02 solution
          is plotted as dash-dotted (short dash-dotted)
          curves.  
} \label{fig:fig5}
\end{figure*}

The changes displayed in Figs.~\ref{fig:fig1} through 
\ref{fig:fig4} cannot be attributed to a single resonance 
or partial wave, as was argued for the GDH observable in 
Ref.~\cite{GDH}.  A number of multipoles show moderate 
changes~\cite{changes}, and only some qualitative 
observations are possible.  For example, a change in the 
$E^{3/2}_{0+}$ multipole can effect the overall shift seen 
in Fig.~\ref{fig:fig1}(d), and changes in either the 
$M_{1+}^{1/2}$ or $M^{3/2}_{1+}$ multipoles produce a 
peaking structure such as is seen in Figs.~\ref{fig:fig3}(b) 
and \ref{fig:fig4}(d).  However, a more complicated change 
in several amplitudes has given the full result.  Some of 
the most prominent changes are displayed in Fig.~
\ref{fig:fig5}.

In summary, we have found it possible to fit recent
double-polarization measurements which initially showed 
significant deviations from both SAID and MAID.  Adding 
these precise new data has actually improved our fit to 
some other quantities, most obviously the 
target-polarization asymmetry, and has been accomplished 
without a substantial degradation of our fit to other 
observables.  The resulting shifts in our multipoles 
indicate that the region above 1~GeV is still 
underconstrained.  Further shifts should be expected 
as the present generation of photon facilities contributes 
data to this region.  In this regard, the proposed 
polarized-target/beam program for Hall~B of JLab would be
extremely useful~\cite{loi}.

\acknowledgments

We thank R.~Gilman for helpful discussions.  This work was 
supported in part by the U.~S. Department of Energy Grant 
DE--FG02--99ER41110.  The authors gratefully acknowledge a 
contract from Jefferson Lab under which this work was done.  
Jefferson Lab is operated by the Southeastern Universities 
Research Association under the U.~S.~Department of Energy 
Contract DE--AC05--84ER40150.

\eject




\begin{thebibliography}{99}

\bibitem{GW_pi}   R.~A.~Arndt, W.~J.~Briscoe, I.~I.~
                  Strakovsky, and R.~L.~Workman, Phys.\ Rev.\ 
                  C\ \textbf{66}, 055213 (2002).
\bibitem{bonn_e2} R.~M.~Davidson, N.~C.~Mukhopadhyay, M.~S.~
                  Pierce, R.~A.~Arndt, I.~I.~Strakovsky, and 
                  R.~L.~Workman, Phys.\ Rev.\ C\ \textbf{59}, 
                  1059 (1999). 
\bibitem{graal}   O.~Bartalini \textit{et al.}, Phys.\ Lett.\ 
                  \textbf{B544}, 112 (2002);
                  J.~Ajaka \textit{et al.}, Phys.\ Lett.\ 
                  \textbf{B475}, 372 (2000).
\bibitem{maid}    Mainz fits are available at the MAID website
                  \hbox{http://www.kph.uni-mainz.de/MAID/}.
                  See also S.~S.~Kamalov, S.~N.~Yang, D.~
                  Drechsel, O.~Hanstein, and L.~Tiator, Phys.\
                  Rev.\ C\ \textbf{64}, 032201 (2001).
                  The Unitary Isobar Model was developed at
                  Mainz, D.~Drechsel, O.~Hanstein, S.~S.~
                  Kamalov, and L.~Tiator, Nucl.\ Phys.\
                  \textbf{A645}, 145 (1999).  MAID refers to
                  the Nov.~2001 version of the MAID solution
                  from S.~Kamalov.
\bibitem{bds}     I.~S.~Barker, A.~Donnachie, and J.~K.~Storrow,
                  Nucl.\ Phys.\ \textbf{B95}, 347 (1975).
\bibitem{GDH}     J.~Ahrens \textit{et al.}, Phys.\ Rev.\ 
                  Lett.\ \textbf{88}, 232002 (2002).
\bibitem{gilman}  K.~Wijesooriya \textit{et al.},
                  Phys.\ Rev.\ C\ \textbf{66}, 034614 (2002).
\bibitem{sabit}   Comparison of SAID and MAID is limited by
                  the MAID energy range, which is expected
                  increase in future versions (S.~Kamalov,
                  private communication).
\bibitem{forced}  R.~L.~Workman, R.~A.~Arndt, and Z.~Li,
                  Phys.\ Rev.\ C\ \textbf{46}, 1546 (1992).
\bibitem{barbour} I.~M.~Barbour, R.~L.~Crawford, and N.~H.~
                  Parsons, Nucl.\ Phys.\ \textbf{B141}, 253 
                  (1978);
                  P.~J.~Bussey \textit{et al.}, Nucl.\ Phys.\ 
                  \textbf{B154}, 492 (1979).
\bibitem{bs79}    P.~J.~Bussey \textit{et al.}, Nucl.\ Phys.\
                  \textbf{B159}, 383 (1979);
                  P.~S.~L.~Booth, DNPL Annual Report, 1977. 
\bibitem{yer}     R.~O.~Avakyan \textit{et al.}, Yad.\ Fiz.\
                  \textbf{53}, 717 (1991) [Sov.\ J.\ Nucl.\
                  Phys. \textbf{53}, 448 (1991)].
\bibitem{said}    The full database and numerous PWAs can
                  be accessed via a ssh/telnet call to the
                  SAID facility \hbox{gwdac.phys.gwu.edu}, 
                  with userid: said (no password), or a link
                  to the website
                  \hbox{http://gwdac.phys.gwu.edu}.
\bibitem{changes} While it is possible to improve the fit by 
                  changing a single multipole, as was found 
                  for the GDH data, this approach tends to 
                  degrade the fit to unpolarized cross section, 
                  as is evident in Fig.~2 of Ref.~\cite{GDH}.
                  This was also mentioned by, R.~Workman, R.~A.~
                  Arndt, and I.~I.~Strakovsky, Phys.\ Rev.\ C\ 
                  \textbf{62}, 048201 (2000).
\bibitem{loi}     R.~A.~Arndt \textit{et al.}, JLab Letter-of-Intent, 
                  LoI--02--101, Newport News, VA, USA, 2002.
\end{thebibliography}
\end{document}